\documentstyle[12pt,epsfig]{article}

\def\frac#1#2{ {{#1} \over {#2} }}
\def\sfrac#1#2{\mbox{\small $\frac{#1}{#2}$}}
\def\half{\mbox{\small $\frac{1}{2}$}}

\def\gtap{\raisebox{-.4ex}{\rlap{$\sim$}} \raisebox{.4ex}{$>$}}
\def\VEV#1{\left\langle #1\right\rangle}

\def\ie{\hbox{\rm i.e. }}

\def\beq{\begin{equation}}
\def\eeq{\end{equation}}
\def\re#1{(\ref{#1})}
\def\L{\Lambda}
\def\b{\beta}
\def\s{\sigma}
\def\g{\gamma}
\def\zs{z_0^{eff}}
\def\as{\alpha_{\sf s}}
\def\Tr{\mbox{Tr}\;}
\def\cren#1{c^{ren}_#1}
\def\cl#1{c^{lat}_#1}
\def\MSbar{\overline{\rm MS}}

\def\cav#1{Cambridge preprint Cavendish--HEP--#1}

\def\np#1#2#3{Nucl.\ Phys.\ B#1 (19#3) #2}
\def\pl#1#2#3{Phys.\ Lett.\ #1B (19#3) #2}
\def\pr#1#2#3{Phys.\ Rev.\ D #1 (19#3) #2}
\def\prep#1#2#3{Phys.\ Rep.\ #1 (19#3) #2}
\def\prl#1#2#3{Phys.\ Rev.\ Lett.\ #1 (19#3) #2}
\def\rmp#1#2#3{Rev.\ Mod.\ Phys.\ #1 (19#3) #2}

\begin{document}
\begin{titlepage}
\renewcommand{\thefootnote}{\fnsymbol{footnote}}
\begin{flushright}
{\sf UPRF-95-418}\\
{\sf IFUM 540-FT}
\end{flushright}
\par \vskip 10mm
\begin{center}
{\Large \bf
Renormalons from eight loop expansion \\
of the gluon condensate  in lattice gauge theory
\footnote{Research supported in part by MURST and by the EC contracts
CHRX-CT92-0051 and CHRX-CT93-0357}
}
\end{center}
\par \vskip 2mm
\begin{center}
        {\bf F.\ Di Renzo and E.\ Onofri} \\
        Dipartimento di Fisica, Universit\`a di Parma and\\
        INFN, Gruppo Collegato di Parma, Italy\\
        \vskip 0.3 cm
        and\\
        \vskip 0.3 cm
        {\bf G.\ Marchesini}\\
        Dipartimento di Fisica, Universit\`a di Milano and\\
        INFN, Sezione di Milano, Italy
\end{center}
\par \vskip 2mm
\begin{center} {\large \bf Abstract} \end{center}
\begin{quote}
We use a numerical method to obtain the weak coupling perturbative
coefficients of local operators with lattice regularization.
Such a method allows us to
extend the perturbative expansions obtained so far
by analytical Feynman diagrams calculations.
In $SU(3)$ lattice gauge theory in four dimensions we compute
the first eight coefficients of the expectation value of
the Wilson loop on the elementary plaquette which is related to
the gluon condensate.
The computed eight coefficients grow with the order much faster than
predicted by the presence of the infrared renormalon associated to
the dimension of the gluon condensate.
However the renormalon behaviour for large order is quite well
reproduced if one considers the expansion coefficients in a new
coupling related to the lattice coupling by large perturbative
corrections. This is expected since the lattice and continuum
$\L$ scales differ by almost two orders of magnitude.
\end{quote}
\end{titlepage}

\section{Introduction}
In asymptotically free theories, such as Yang-Mills gauge theories
in four dimensions, the short distance distributions are
reliably approximated by the perturbative expansion in $\as(Q)$,
the running coupling at the large momentum scale $Q$.
However, even in this case the perturbative expansions are affected
\cite{Borel}-\cite{G} by ambiguities due to infrared (IR) renormalons,
\ie singularities on the integration contour in the Borel variable
conjugate to $\as(Q)$.
The origin of these singularities is based on the renormalization group
properties. Since in computing short distance distributions also small
momenta $k$ are involved in loop integrations, one has singularities due
to the Landau pole in the running coupling $\as(k)$ at $k \ll Q$.

 The IR renormalons in the expectation value of composite operators have been
studied in \cite{ON}
in the case of the $O(N)$ non-linear sigma model for $N \to \infty$.
 One finds that the related ambiguity is absorbed
by non-perturbative contributions present in the various terms of the
operator product expansion.
A similar mechanism should also hold  in the Yang-Mills case and
prescriptions for resolving the ambiguities have been proposed
\cite{M,G}.

Perturbative expansions of short distance distributions in QCD,
such as the total $e^+e^-$ hadronic cross section and total decay
widths, are known \cite{3loops} up to three loops in the $\MSbar$
renormalization prescription.
It is difficult to expect that one may reliably explore
 the properties of the IR renormalons on the basis of
 these not very long expansions.

Recently we have proposed \cite{DMMO} a numerical method to obtain
 the weak coupling perturbative expansion of local observables
in pure Yang-Mills gauge theory in four dimensions in the lattice
regularization.
The method is based on the idea of
taking the weak coupling expansion in the
Langevin stochastic formulation of the theory \cite{PW} and  solving
numerically the truncated set of equations corresponding to a given
perturbative order.
By this method we hoped to be able to obtain perturbative
expansions longer than the ones obtain by analytical methods
\cite{3loops,Pisa}.

In this paper we present the results obtained in $SU(3)$ lattice gauge
theory in four dimensions for the first eight coefficients of the
expectation value of the ``plaquette variable'', \ie the Wilson loop on the
elementary plaquette of lattice size $a$.
The first four coefficients have been presented in Ref.~\cite{DMMO}.
On the basis of
 these new results we try to study the asymptotic behaviour of the
coefficients for large orders and we try to analyze the IR renormalon
singularities.

We consider the standard $SU(3)$ gauge lattice action
\beq\label{S}
S[U] = - \frac{\b}{6}\sum_{P} \Tr (U_P+U^{\dagger}_P)
\,,
\;\;\;\;\;\;\;\;\;\;
\frac \b 6 = \frac 1 {4\pi \as}
\,,
\eeq
where $\as$ is the usual  coupling at the lattice scale.
The sum extends to all plaquettes $P$ in a hypercubic lattice in four
dimensions, $U_P$ is the plaquette gauge field obtained from the link
variable $U_\mu(x)=\exp \{ A_\mu(x)/\sqrt \b \}$.
The observable we consider is the dimensionless expectation value
of the plaquette
\beq\label{plaq}
W^{lat}_4(\b) \equiv 1- \sfrac{ 1}{3} \VEV{\Tr U_P}
\, .
\eeq
and we compute the coefficients $\cl n $ of the weak coupling
expansion \beq\label{pert}
W^{lat}_4(\b) \sim \sum_{n=1} \cl n  \b ^{-n}
\,,
\eeq
The physical interest of $W_4(\b)$ is that, in the operator product
expansion of this observable, the first non trivial term is the gluon
condensate $\VEV{\as F^2}$ which has dimension four. This quantity
is renormalization group invariant and therefore its contribution to
$W_4(\b)$ is proportional to $(a\L)^4$, where $\L$ is the Yang-Mills
scale given, at two loops, by
\beq\label{aL}
(a\L)^2=C\,\left(\frac{\b}{6b_0}\right)^{b_1/b_0^2}\; e^{-\b/6b_0}\,,
\;\;\;\;\;\;
b_0=11/(4\pi)^2\,,
\;\;\;\;\;\;
b_1=102/(4\pi)^4
\,.
\eeq
There is no perturbative expansion for the quantity $a\L$
and therefore also for the physical condensate $\VEV{\as\,F^2}$.
The only perturbative contribution in \re{pert} is given by the first
trivial term in the operator product expansion of $W^{lat}_4(\b)$,
\ie the unit operator.
The coefficients $\cl n$ are obtained from quartically divergent Feynman
diagrams in which the lattice size $a$ plays the r\^ole of the UV cutoff.
The IR renormalons are present in this perturbative expansion since
the momenta along the loops are integrated down to zero.

The result of our calculation is that the first eight coefficients
$\cl n$ grow much faster then expected from the conventional
analysis of IR renormalons.
For this comparison one uses the expansion
\beq\label{pertr}
W_4^{ren}(\b)= \sum_{n=1} \cren n \b^{-n}
\eeq
with $\cren n$ determined by the presence of an IR renormalon
singularity associated to the gluon condensate, \ie of dimension four.
We find then that $\cl n$ grow with $n$ much faster than $\cren n$ for
$n \le 8$.
This feature was already observed in the four loop calculation
\cite{DMMO} but now it is much more clear. Such a fast growth could
be due to large subleading contributions in $n$.
Indeed, as well known \cite{scale}, the lattice and continuum
$\L$ scales differ by almost two order of magnitude so that
the perturbative relation between the lattice and continuum couplings
\beq\label{r}
\b_{cont} = \b_{latt} -r +{\cal O}(\b_{latt}^{-1})
\,,
\eeq
involves the large correction $r = 12 b_0 \ln (\L/\pi\L_{latt})$.
This suggests to compare the lattice expansion
$W_4^{lat}(\b_{latt})$ with the renormalon expansion
$W_4^{ren}(\b_{cont})$ in eq.~\re{pertr}
where the two couplings are related by \re{r}.
We find that for $r\sim 2.4$
the coefficients of $\b_{latt}^{-n}$ in the expansion of
$W_4^{ren}(\b_{cont})$ agree quite well with the coefficients
$\cl n$ for $n \gtap 4$.
Such a large value for $r$ agrees with the values obtained
\cite{scale} in usual continuum renormalization schemes although it
appears somehow larger than the one of $\MSbar$ renormalization.

In Sect.~2 we briefly recall the method for obtaining numerically the
perturbative coefficients $c^{LGT}_n$ for the $SU(3)$ lattice gauge
theory in four dimension.
In Sect.~3 we present our results and discuss the statistical
significance.
In Sect.~4 we recall the asymptotic behaviour of the coefficients
$c^{LGT}_n$ deduced from the usual IR renormalon analysis by using the
one and two loop running coupling.
In Sect.~5 we discuss the behaviour of the eight computed coefficients
in comparison with the one predicted by the IR renormalon
analysis including the two loop correction of the running coupling.
Sect.~6 contains some final comments.
We include in the Appendix a conjecture, not plausible on the
light of the present analysis, that the fast growth
of the lattice coefficients $\cl n$ could be due, at least partially,
to additional infrared singularities in the Feynman diagrams.

\section{Weak coupling expansion on the computer}
Let us briefly recall the essential ingredients of our method
for calculating the weak coupling expansion in the lattice
regularization.
First we describe the method for a scalar field; then we proceed to the
case of $SU(N)$ lattice gauge theory and its gauge fixing problem.
(For a recent review about Langevin simulations see {\em e.g.}
\cite{Kronfeld}).

The Langevin method
consists in the simulation of a stochastic dynamical system having the field
configuration space as its state space. Time evolution is
dictated by the general equation
\begin{equation}\label{eq:parwu}
 \frac{\partial\phi(x,t)}{\partial t} = - \frac{\partial S[\phi]}
 {\partial \phi(x,t)} + \eta(x,t), \
\end{equation}
where $\phi$ is the field, $S[\phi]$ the action and $\eta$ a
Gaussian random noise satisfying the normalization
\begin{equation}
 \VEV{\eta(x,t)\eta(x^{'},t^{'})} = \delta(x-x^{'})\delta(t-t^{'}). \
\end{equation}
As a matter of fact, stochastic dynamics is devised in such a way that
time averages along a trajectory
converge to averages with respect to the Gibbs measure
\begin{equation}
 \frac{1}{T}\int_{0}^{T} dt\; \VEV{O[\phi(t)]}_{\eta} \to
 \frac{1}{Z}\int D\phi\; O[\phi]\; e^{-S[\phi]}.
\end{equation}
where the suffix $\eta$ denotes an average over the Gaussian noise.

In order to implement Eq.\ref{eq:parwu} on a computer,
one can take $t$ discrete with a time step $dt=\varepsilon$ \cite{Parisi2}:
\begin{equation}
 \phi(x,n+1) = \phi(x,n) - f(x,n),
\end{equation}
where
\begin{equation}
 f(x,n)=\varepsilon  \frac{\partial S}{\partial\phi(x,n)}+\sqrt{\varepsilon}
 \eta(x,n)
\end{equation}
and now $\eta$ is normalized by:
\begin{equation}
 \VEV{\eta(x,n)\eta(x',n')} =\delta_{x x'}\delta_{n n'}.
\end{equation}
In this discrete form, Langevin equation is affected by a systematic error,
which makes it necessary to extrapolate the results at $\varepsilon\to 0$
or to devise some higher order approximation to the continuum
stochastic equation. In this paper we adhere to the simple recipe
which consists in performing the simulation at several values
of $\varepsilon$ with a linear fitting to $\varepsilon=0$.

Let us now assume that the action is given by a free part (say
$S_0=\half\sum_x \phi(x)(-\Delta+m_0^2)\phi(x)$) plus
an interaction term $g\,\sum_x \phi^4$;  the solution
to Langevin's equation will then depend
parametrically on the coupling constant $g$; we
insert the formal expansion
\begin{equation}\label{eq:expansion}
\phi(x,t,g) \sim \sum_{k\ge 0} g^k \phi^{(k)} (x,t)
\end{equation}
into the Langevin evolution equation to obtain a hierarchy
of stochastic evolution steps of the kind
\begin{eqnarray*}
\phi^{(0)}(x,n+1)  &=& \phi^{(0)}(x,n) - \varepsilon (-\Delta+m_0^2)
\phi^{(0)}(x,n) + \sqrt\varepsilon \eta(x,n)\\
\phi^{(1)}(x,n+1) &=& \phi^{(1)}(x,n) - \varepsilon (-\Delta+m_0^2)
\phi^{(1)}(x,n)
        - \varepsilon \left(\phi^{(0)}(x,n)\right)^3\\
\phi^{(2)}(x,n+1) &=&  \phi^{(2)}(x,n) - \varepsilon (-\Delta+m_0^2)
\phi^{(2)}(x,n)
        - \varepsilon \left(\phi^{(0)}(x,n)\right)^2\,\phi^{(1)}(x,n)\\
\ldots
\end{eqnarray*}

If we want to measure  the expansion in $g$ of
any given observable $O[\phi]$ we have just to insert the formal
expansion \re{eq:expansion} and take averages on the random
process. For instance the expansion for the
energy density will be given by
$$
\VEV{{\cal E}[\phi]} = \VEV{(\nabla \phi(x)) ^2}
\sim
\VEV{(\nabla\phi^{(0)})^2}
+ 2 g \VEV{\nabla\phi^{(0)}\cdot \nabla\phi^{(1)}}
+g^2 \VEV{(\nabla\phi^{(1)})^2
+2\nabla\phi^{(0)}\cdot\nabla\phi^{(2)}}
+\cdots
$$
The idea is very general and it can be applied to spin systems or
to lattice gauge models.

In $SU(3)$ lattice gauge  theory the action is given by Eq.\re{S}
defined on a four dimensional periodic lattice.
The link gauge variables $U_{\mu}(x)$ are $SU(N)$ matrices labelled
by the vector index $\mu=1,\ldots 4$ and by the
site $x$  of the lattice. Each configuration is then
described by $36$ complex numbers at each lattice site.
In this case the Langevin evolution step in discrete time $\varepsilon$
is given (see Ref~\cite{Bat})
\begin{equation}
 U_\mu(x,n+1) =  e^{-F_\mu(x,n)} U_\mu(x,n),
\end{equation}
where
\begin{equation}
 F_\mu(x)  = \frac{\varepsilon\beta}{4N} \left[ \sum_{U_P\supset U_{\mu}}
(U_P-U_P^{\dagger})  - \frac{1}{N}
\sum_{U_P\supset U_{\mu}}\Tr(U_P-U_P^{\dagger}) \right]
+ \sqrt\varepsilon H_\mu(x)
\end{equation}
and $H_\mu(x)$ is a traceless antihermitian  matrix extracted at random
independently at each step from a Gaussian ensemble with normalization
given by
\begin{equation}
 \VEV{H_{ik}(x,n)\overline{H}_{lm}(x',n')}_H = [\delta_{il}\delta_{km}-
 \frac{1}{N}\delta_{ik}\delta_{lm}]\delta_{xx'}\delta_{nn'}.
\end{equation}
In the weak coupling limit we can parameterize the link variables
in terms of the usual gauge potentials $A_\mu$, namely we set
\begin{equation}
\begin{array}{rcl}
 U_{\mu}(x)=e^{A_\mu(x)/\surd\beta},
&A_\mu(x)^{\dagger}
=-A_\mu(x),
&\Tr A_{\mu}=0.
\end{array}
\end{equation}
It follows that Langevin's equation takes the form:
\begin{equation}
 e^{A_\mu(x,n+1)/\surd\beta} =  e^{-F_\mu(x,n)}e^{A_\mu(x,n)/\surd\beta}\,,
\end{equation}
where $n$ is Langevin's discrete time.
As in the scalar case the trajectories
$A_{\mu}(x,n)$ will depend parametrically on $\beta$:
we  expand them  in a formal series in powers of  $\beta^{-1/2}$
\begin{equation}
 A_\mu (x) =\sum_{k\ge 0 }\beta^{-k/2} A_\mu^{(k)}(x)\,.
\end{equation}
and we  rescale the time step
$\varepsilon=\tau/\beta$ in such a way that
 Langevin's algorithm can be reformulated
in terms of the fields $A_\mu^{(k)}$. The formulae rapidly get
rather cumbersome but they can  easily be handled by a symbolic
language or defining convenient structures as it is natural
within the {\sf ZZ} language of APE.

We have reported the calculation of the average plaquette
up to fourth loop ($\beta^{-4}$) in \cite{DMMO}.
It was shown that in order to obtain a stable process one
has to impose some kind of gauge fixing; this fact seems quite
natural in view of the infinite number of zero modes
which one has to deal with in the expansion around a
given classical vacuum. A gauge fixing  procedure for
Langevin quantization was proposed long ago by Zwanziger \cite{Zwanz}
 and it was implemented on the lattice in
\cite{Pietro}.
The idea consists in introducing a  new source in
Langevin's equation which does not modify the asymptotic
stationary distribution $\exp\{-S\}$ but it acts as an attractor
towards the Landau gauge manifold defined by $\nabla\cdot A_\mu=0$.
The implementation on the lattice consists in a gauge transformation
which is executed after each Langevin step:
\begin{equation}
 U_\mu(x) \rightarrow  e^{w[U_{\mu}]} U_\mu(x)\; e^{-w[U_{\mu}(x+\mu)]},
\end{equation}
with
\begin{equation}
 w[U_{\mu}(x)]  = \lambda\sum_{\mu}\left( \Delta_{-\mu}\left[ U_{\mu}(x)-
 U_{\mu}^{\dagger}(x-\mu)\right]\right)_{\rm{tr}},\;\;\;\;\;
\Delta_{-\mu} U_{\nu}(x) \equiv U_{\nu}(x)-U_{\nu}(x-\mu),
\end{equation}
where the suffix ``tr'' denotes the traceless part of the matrix
 and $\lambda$ is a free parameter which
we choose proportional to the time step $\tau$.

\section{ Eight loop results for the plaquette}
The algorithmic setup presented in the last section lends itself
to the possibility of performing high order weak coupling expansions;
needless to say, since
the calculation to $n-th$ loop involves $2n-1$ auxiliary
fields $A^{(0)}, A^{(1)}, \ldots, A^{(2n-2)}$, it tends to be
quite demanding in terms of computer memory and cpu time.
The results reported  here were obtained on an array processor
with 64 units of the APE family. They required approximately
600 hours (roughly equivalent to 5000 hours
on a single cpu of a Cray YMP)
for the final run, which consisted of 24 independent
Langevin histories of an average 4000 steps. Since the finite size effects
on the single plaquette expectation value are rather small,
we work on a $8^4$ lattice with periodic boundary conditions which
fits just fine on the memory of a single APE board, thus
obtaining eight statistically independent Langevin histories
at each run. There were three independent runs
at $\tau=.02,\;.015,\; .01$ from which we extrapolate at the
desired $\tau=0 $ limit. We report in Fig.~1
the signal we get for $\tau=0.01$. Since the higher loop
observables depend on the lowest orders, we chose to start
measuring the various observables in cascade, in order to let
the system gradually thermalize.

\begin{figure}[t]\label{fig:histories}
\caption{The eight independent time averages for each of the
observables yielding an estimator of $c_1^{LGT},\ldots,c_6^{LGT}$
at $\tau=0.01$.}
\end{figure}

The statistical errors are computed
by taking the standard deviation divided by square root of the number
of statistically independent samples,
which we estimate to be around 64. This is a conservative estimate
based on the presence of eight independent Langevin trajectories
assuming an autocorrelation time of 300 steps.
We plan to have a reliable
measure of autocorrelation in the near future, when longer
Langevin histories will be available. Fig.~2
reports the process of extrapolation, together with the central
value and the estimated statistical error for the first {\em eight}
expansion coefficients.

\begin{table}
\begin{center}
\begin{tabular}{|c||c|c|c|c|c|c|c|c|}\hline
$n$        & 1   &  2  &  3  &  4 &  5 &  6 & 7 & 8  \\\hline
$c_{\rm Langevin}$
      & 1.998(1) &  1.218(1) & 2.940(5) & 9.28(2) & 34.0(2) & 134.9(9)& 563(5)
& 2488(29)\\
$c_{\rm analytic}$
      & 2        & 1.218(7)  & 2.9602 & --- & --- & --- & --- & --- \\\hline
\end{tabular}
\end{center}
\vskip .1 in
\caption{The expansion coefficients $c_n^{LGT}$
obtained by the stochastic method and the known coefficients
from Ref.(9).
}
\end{table}

\begin{figure}[t]\label{fig:Histograms}
\caption{The extrapolation at $\tau=0$
of the first eight perturbative
coefficients $c_n^{LGT}$ on the $8^4$ lattice. }
\end{figure}

We end this section by a remark about the finite size
corrections. To get a rough estimate on them we performed
a series of runs on a very small lattice ($L^4$ with $L=4$) and
on the largest feasible lattice on our APE ($L=12$).
We observe an increase in the values of $c_n(L)$ with $L$ which is
more sensible for higher loops. Essentially is appears rather
reasonable that finite size corrections become important
as $n/L \sim 1$. The overall picture seems to agree with
the analysis of Ref.\cite{Pisa}, but  a deeper analysis
would admittedly be desirable.

\section{ IR renormalon analysis}
We recall here the usual conventional IR renormalon analysis.
The Feynman diagrams of the plaquette expectation value $W_4(\b)$
are quadratically divergent.
In general the perturbative contribution $W_{2\s}(\b)$ to the expectation
of a composite operator which has UV divergences of
degree $2\s$ (for the plaquette $2\s=4$) has the form
\beq\label{Ws}
W_{2\s}(\b) = \int_0^{Q^2}\frac{dk^2}{k^2}\,
\left( \frac{k^2}{Q^2}\right)^{\s} \, f(k) \,, \;\;\;\;\;\;\;
\frac{6}{\b}=4 \pi \as(Q)\,,
\eeq
where $Q$ is of the order of the inverse of lattice size,
$Q \sim 1/a$, which plays the r\^ole of the UV cutoff.
The function $f(k)$ is dimensionless, renormalization group invariant
and to first order is given
by the one gluon exchange diagram, \ie is proportional to $\as(Q)$.
In QED the contributions from the resummation of fermion loop diagrams
\cite{Borel} gives $f(k)=\alpha(k)$, the running coupling in QED.
The important simplification is that, from Ward identities, the vertex
and wave function corrections cancel.
In QCD the situation is different. Actually  to reconstruct $\as(k)$
one needs also vertex and wave function corrections which do not cancel.
So in QCD one needs to consider more diagrams and there are no
simple criteria to select the important ones.
By taking into account that $f(k)$ is renormalization group invariant
one assumes that also in Yang-Mills gauge theory
\beq\label{fk}
f(k)=\as(k)
\,.
\eeq
In this case the Landau pole in $\as(k)$ is at a small
value $k \ll Q$ and therefore the integral in \re{Ws} diverges.
The perturbative expansion of $W_{2\s}(\b)$ in $\as(Q)$  is obtained
by expanding $\as(k)$. The Landau pole disappears and each
coefficient is finite.
The divergence of \re{Ws} at small $k$ is reflected on the fact that
the coefficients grow fast and the expansion
diverges. A nice framework for discussing this feature is given by
the Borel transform.
Introducing the (Borel) variable $z$
\beq\label{zdef}
z \equiv z_0\left(1-\frac{\as(Q)}{\as(k)}\right)
\,,
\;\;\;\;\;\;
z_0=\frac{\s}{6b_0}
\,,
\eeq
and using \re{aL}
the integrand of \re{Ws} can be written
\beq
\frac{dk^2}{k^2}
\left(\frac{k^2}{Q^2}\right)^{\s}\,\as(k)
=
\frac{4 \pi b_0 \as^2(k)}{\b(\as(k))}
\;
\frac{6}{4 \pi \s}\,
dz \, e^{-\b z}\,\left( 1-\frac{z}{z_0}\right)^{-1-\g} \,,
\;\;\;\;\;\;
\g=\s\, \frac{b_1}{b_0^2}
\,,
\eeq
where $\b(\as)$ is the beta function
\beq\label{jk?}
\frac{\b(\as(k))}{4\pi b_0\as(k)^2}=
1 + 4\pi b_0\frac {b_1}{b_0}\as(k) + \cdots
\eeq
We have now to consider the range of integration for $z$
corresponding to $0 <k< Q$.
Beyond the Landau pole at $z=z_0$ the mapping \re{zdef}  becomes
ambiguous. However at two loops we have
$$
z=\frac{4 \pi \s}{6}\,
\as(Q)\, \left\{1+4\pi \frac{b_1}{b_0}\,\as(Q) \right\}
\, \ln \frac{Q^2}{k^2}
+\cdots
\,,
$$
thus we assume the range $0 < z < \infty$.
Neglecting the corrections in \re{jk?} we obtain
the Borel representation
\beq\label{Bint}
W^{ren}_{2\s}(\b) = \int_0^{Q^2}\frac{dk^2}{k^2}\,
\left( \frac{k^2}{Q^2}\right)^{\s} \, \as(k)
\,=\,
\frac{6}{4\pi \s} \int_0^\infty dz
e^{-\b z}\left(1-\frac{z}{z_0}\right)^{-1-\g}
\,,
\eeq
with the perturbative coefficients given by
\beq\label{Bexp}
W^{ren}_{2\s}(\b)
\;=\;
\sum_{n=1} \,\cren n (2\s,b_1) \; \b^{-n}
\,,
\;\;\;\;\;\;\;
\cren n (2\s,b_1) \;=\; C \; \frac{\Gamma(\g+n)}{z_0^{n}}
\,,
\;\;\;\;\;
\g = \s\, \frac{b_1}{b_0^2}
\,,
\;\;\;\;\;
z_0=\frac{\s}{6b_0}
\,.
\eeq
where $C$ is a numerical constant.
A correction in the integrand in \re{Ws} proportional to
$\as(k)$, as given for instance by taking into account the two
loop contribution of the beta function in \re{jk?}, simply modifies
the normalization constant $C$.
The IR renormalon singularity on the integration contour
at $z=z_0$ corresponds to the Landau pole of $\as(k)$.
For the plaquette we have
\beq\label{z0}
2\s=4 \,,
\;\;\;\;\;\;\;
z_0=\frac{1}{3b_0}= 4.785...
\eeq
To give a meaning to the representation \re{Bint} one has to
give a prescription to specify the way the integration contour
avoids the singularity $z=z_0$.
The various prescriptions differ by contributions
proportional to the residue at the pole. The ambiguity in \re{Bint}
is then given by
\beq\label{amb}
\Delta W_s(\b) \sim e^{-\b z_0}=e^{-\frac{\b\s}{6b_0}}\sim (a\L)^{2\s}
\,.
\eeq
For the plaquette $2\s=4$ this ambiguity is of the same order as
the condensate ($\VEV{\as F^2} \sim \L^4$), \ie the first non trivial
term in the operator product expansion of the plaquette.
This suggests that the IR renormalon ambiguity should be absorbed
by non perturbative contributions coming from the various terms of the
operator product expansion.

\section{Renormalon and the numerical coefficients}
In Fig.~3 we plot the eight coefficients $\cl n$ of the
perturbative expansion of the plaquette $W^{lat}_4(\b)$
in \re{pert} obtained from the numerical method reported in Sect.~3.
We plot also the coefficients $\cren n$ for the IR renormalon in
\re{Bexp} with $2\s=4$ and $z_0$ is given in \re{z0}.
The normalization constant $C$ is fixed in such a way that
$\cren 8=\cl 8$.
{}From the comparison we conclude that the growth of the
coefficients $\cl n$ is much stronger than the one of
$\cren n (4,b_1)$.

A simple explanation for the discrepancy observed between $\cren n$
and $\cl n$ is that the coupling $\b=\b_{latt}$ in the lattice action
\re{S} and the coupling suitable for the analysis of the IR renormalons
in \re{Ws} are related by large perturbative corrections.
One expects that the continuum coupling $\b_{cont}$ is the suitable
coupling for the renormalon analysis.
As well known \cite{scale} in the Yang-Mills case the $\L$ scales
in the lattice
and continuum renormalization schemes are quite different
\beq\label {K}
\L_{cont}\;=\; K \, \L_{latt}
\,,
\eeq
with large values for $K$.
In the $\MSbar$ scheme one has $K=28.8$. If the coupling $\as(Q)$
is defined by the static potential between quark-antiquark pair
or by the three point vertex one finds the constants
$K=46.1$ or $K=69.4$ respectively.
This relation can be converted in terms of a relation between the
lattice and continuum scale.
{}From \re{aL} and \re{K} one finds the relation \re{r} between the two
couplings where $r=1.85$, $r=2.25$ and $r=2.6$ in the renormalization
scheme in
which the coupling is defined in $\MSbar$, by the static potential or
by the three point vertex respectively.

In order to analyze the renormalon behaviour of the computed lattice
coefficients $\cl n$ we compare the lattice expansion \re{pert}
in the coupling $\b_{latt}^{-1}$ with the renormalon expansion
\re{Bexp} in the coupling $\b_{cont}^{-1}$
\beq\label{Bexpr}
W^{ren}_{2\s}(\b_{cont})
= \sum_{n=1} \,\cren n (2\s,b_1) \; \b_{cont}^{-n}
= \sum_{n=1} \,C_n^{ren}(2\s,b_1) \; \b_{latt}^{-n}
\,,
\eeq
where we assume the following relation between the two couplings
\beq\label{rr}
\b_{cont}= \b_{latt} - r - \frac{r'}{\b_{latt}}
\,,
\eeq
so that $C_n^{ren}(2\s,b_1)= \cren n (2\s,b_1)$ for $r=r'=0$.
The parameter $r'$ takes contributions from three loops. We have
$r'=6rb_1/b_0 + 36\delta b_2/b_0$ with $\delta  b_2$ the difference
between the three loop beta function coefficients in the lattice
and continuum renormalization schemes.

In Fig.~3 we plot the new coefficients $C_n^{ren}(4,b_1)$
obtained by fitting the two parameters $r$ and $r'$. We find
$r=2.41$ \footnote{Notice that this value is remarkably close
to Lepage and Makenzie's (\cite{scale}) taking into account the
effective momentum $q^*\sim 2.8$ proper to $\alpha_V$.}
and $\delta b_2=.12$.
For $n \gtap 4$ we have a very good agreement between
$\cl n(4,b_1)$ and $C_n^{ren}(4,b_1)$.

The value of $r=2.41$ corresponds to the scale constant in
\re{K} given by $K=56.1$ which is between the ones obtained by
defining $\as(Q)$ by the static potential or by the three vector
coupling and smaller than the one in the $\MSbar$ scheme.

Since the value of $r$ is large, the relation between
$\cl n$ and $C_n^{ren}$ involves large cancelations. This implies
that the results are sensible to $r$, as is rather evident from
Fig~3.

We have studied also the dependence on the two loop corrections
by setting $b_1=0$ in the coefficients $C_n^{ren}(4,b_1=0)$.
In order to obtain the fit with the lattice coefficients for
$n \gtap 4$ we find $r=4.13$ rather larger than the two loop value
given before.

\begin{figure}\label{fig3}
\caption{
The values of $\cl n,\, (n=1\ldots,8)$ obtained
by the stochastic method together with renormalon coefficients
$\cren n(4,b_1)$   and
$\bar \cren n (4,b_1,r)$ with $r=2.41$.
}
\end{figure}

\section{Final considerations}
We used a numerical method to obtain on a computer long perturbative
expansion in four dimensional Yang-Mills theory with a lattice
regularization. The expansion parameter is the lattice coupling
$\b_{latt}=2\pi \as/3$ entering in the Wilson action \re{S} at the
lattice size $a$.
The large order behaviour of the plaquette perturbative
coefficients $\cl n$ agrees for $n \gtap 4$ with the presence of a
IR renormalon at \re{z0} with $2\s=4$ corresponding to the dimension
of the gluon condensate $\VEV{\as \, F^2}$.
However to obtain such a result one has to consider large corrections
in the perturbative relation \re{r} between the lattice coupling
$\b_{latt}$ and the continuum coupling $\b_{cont}$ in which one discuss
the renormalons.
Such a large perturbative correction ($r=2.41$) is expected
since, as well known \cite{scale}, the $\L$ scales for the lattice and
continuum regularizations differ almost by two orders of magnitude.
It should be noticed that since $\cl n$ grow fast with $n$ and $r$ is
large the coefficients in the continuum regularization are
obtained in terms of the lattice ones by large cancelations.

Since there are large perturbative corrections between lattice and
continuum couplings it may be important to know the perturbative
relation in \re{r} with many more coefficients.
This could be attempted for instance by computing perturbatively
by our method the static quark-antiquark potential.

\begin{center}
Acknowledgments
\end{center}
All the numerical results presented here have been obtained
on a APE computer; we are indebted to Nicola Cabibbo, Franco
Marzano and the APE group for providing us with this
facility.

\vskip 1. cm
\par\noindent
{\bf Appendix}

As discussed in the text the plausible explanation of the quite
fast growth of
the lattice coefficients $\cl n$ is the large perturbative
corrections between the lattice and continuum couplings
\re{r}. However we would like to add an alternative explanation.
The fast growth could be explained by an effective value of the
dimension $2\s$ much smaller than the actual value $2\s=4$.
Indeed for the effective value of $2\s=1$ the coefficients
$\cren n (1,b_1)$ in \re{Bexp} fit the numerical values of $\cl n$
for $n \gtap 3-4$.
To explain this smaller effective dimension
suppose that, as suggested by perturbative QCD studies
(see \cite{IR,ABCMV}), the argument of the running coupling $\as$
is modified by infrared corrections of Feynman diagrams in the Minkowski
form so that the integrand $f(k)$ is \re{fk} is given by
\beq\label{HYP}
f(k)=\int_0^1 dx \as(xk)
\,.
\eeq
The perturbative contributions of $f(k)$ in the expansion parameter
$\as(k)$ are finite
and given by integrations over $\ln ^n x$. Resumming the expansion in
$\as(k)$ one has
a Landau pole in the variable $xk$ instead of $k$.
By assuming for the integrand in \re{Ws} the expression in \re{HYP}
one finds that the renormalon contribution in \re{Ws} is
\beq\label{Ws1}
W^{ren}_{2\s}(\b)
=\int_0^{Q^2} \frac{dk^2}{k^2}\,
\left( \frac{k^2}{Q^2}\right)^{\s} \,\int_0^1 dx\,\as(xk)
=\frac {1}{2\s-1} \int_0^{Q^2} \frac{dq^2}{q^2}
\,\as(q)
\,\left\{ \left(\frac{q^2}{Q^2}\right)^{\half}
         -\left(\frac{q^2}{Q^2}\right)^{\s}  \right\}
\,.
\eeq
The first term, which gives the dominant contribution, is just the
expression in \re{Ws} with dimension $2\s \to 1$ with the Borel
singularity at the universal value independent of the dimension $\s$
\beq\label{eff}
2\s^{eff}=1 \,. \;\;\;\;\;\ \zs=\frac{1}{12 b_0} = 1.196 \cdots
\eeq
This Landau singularity
would correspond to a perturbative ambiguity given by
\beq\label{amb1}
\Delta W_2(\b) \sim e^{-\b \zs} \simeq (a\L)
\,.
\eeq
This large non perturbative corrections has been observed for
quantities such as the average trust \cite{W}.

The prove of the conjecture \re{HYP} would need a difficult analysis
since one should study contributions in Feynman diagrams
of terms with infrared singularities which are integrable.
If the contributions in \re{HYP} are present one should care about
the mechanism to resolve the perturbative ambiguity in \re{amb1}.
There are no operators to cure the ambiguity
associated to dimension $2\s_{eff}=1$.
There is no contradiction with the analysis in the $O(N)$
non-linear planar sigma model in two-dimensional \cite{ON}
in which the position of the renormalon is actually related to the
real dimension $2\s$.
In two dimensions the structure of Feynman diagram singularities is
completely different from the case of Yang-Mills theory in
four dimensions especially if one uses the Minkowski form.
The only connection between these models is the sign of the beta
function, \ie the $\as^n(Q)\,\ln^n k$  resummation.

\end{document}